\documentclass[%
 reprint,
 amsmath,amssymb,
 aps,
]{revtex4-2}

\usepackage{graphicx}
\usepackage{xcolor}
\usepackage{dcolumn}
\usepackage{bm}

\usepackage{soul}
\sethlcolor{yellow}

\begin{document}

\preprint{APS/123-QED}

\title{Exact Solutions to Acoustoelectric Interactions in Arbitrary Geometries}

\author{William W. Roberts}
\affiliation{Electrical, Computer, and Energy Engineering, University of Colorado Boulder}

\author{Matt Eichenfield}
 \email{matt.eichenfield@colorado.edu}
\affiliation{Electrical, Computer, and Energy Engineering, University of Colorado Boulder}%
\affiliation{Sandia National Laboratories}
\date{\today}

\begin{abstract}
Acoustoelectric interactions occur when free carriers in a semiconductor interact with the fields of an acoustic wave in a piezoelectric medium. These interactions can amplify acoustic waves, as well as give rise to extremely large phononic nonlinearities and strong non-reciprocal effects. The field of acoustoelectric devices is currently dependent on analytical and perturbative solutions for the two simplest arrangements of piezoelectric-semiconductor materials. While these canonical models have allowed the field to advance substantially, new geometries are arising that do not satisfy assumptions integral to these models. These assumptions include the treatment of the interactions between the acoustic fields and free carriers as weak, the neglect of the tensorial nature of the material properties, the omission of the spatial variations in the phonons' electric field profiles, and the disregard of elastic coupling across material boundaries, among others. We develop, for the first time, a finite element method (FEM) model to solve for acoustoelectric interactions in arbitrary geometries that avoids making the assumptions of the canonical models. We verify the FEM model using results for amplification, dispersion, and non-reciprocity obtained from the canonical models in their regime of validity. We then examine the acoustoelectric effect in two geometries not covered by the canonical models: a thin piezoelectric film placed on a semiconductor substrate and a fully 2D waveguide under a thin semiconductor layer. This work lays the foundation for accurate modeling of arbitrary acoustoelectric geometries such as those currently being developed for all-acoustic radio frequency (RF) signal processing, acoustoelectrically enhanced photonic devices, and quantum acoustoelectric devices.
\begin{description}
\item[Keywords]
Finite Element Method, Acoustoelectrics, Integrated Phononics, Non-Hermitian Physics 
\end{description}
\end{abstract}

\maketitle


\section{Introduction}

\begin{figure}
\includegraphics[width=0.95\columnwidth]{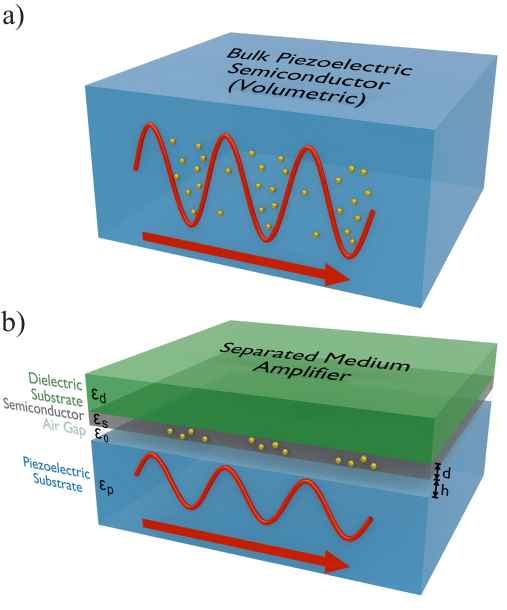}
\caption{a) Bulk piezoelectric semiconductor where analytical models for the acoustoelectric effect have been derived for plane waves. b) Separated medium heterostructure with an air gap between the piezoelectric medium and the semiconductor designed of Rayleigh wave amplification where a perturbation theory has been developed to model the acoustoelectric effect.} 
\label{Fig:Intro}
\end{figure}

The acoustoelectric effect arises from the interactions between the fields of an acoustic wave in a piezoelectric medium and the free carriers inside of a semiconductor \cite{White1967}. These interactions dramatically change the acoustic properties of a material resulting in large shifts in wave velocity \cite{Blotekjaer1964,White1962}, attenuation and gain \cite{Hackett2021,malocha2021}, non-reciprocity \cite{Hackett2023,mansoorzare2020,mansoorzare2021}, giant effective phononic nonlinearities \cite{Conwell1971,Hackett2024}, and acoustoelectrically enhanced optomechanical interactions \cite{Otterstrom2023}. Approximate but accurate expressions for these effects have been derived in the case of bulk piezoelectric semiconductors, where the material platform is homogeneous \cite{Hutson1962,Conwell1971}. Additionally, a perturbation-based model has been derived for the separated medium amplifier, where the moving charge is separated from the piezoelectric material by a small air gap and coupled to the acoustic wave via the electric field \cite{Lakin1969_1,Kino1971}. The geometries for these two canonical models are shown in Figures \ref{Fig:Intro}a and \ref{Fig:Intro}b, respectively. Despite the fact that these canonical models have allowed accurate predictions of acoustoelectric devices for amplification \cite{Lakin1969}, radiofrequency (RF) switching \cite{Storey2021}, nonlinear mixing \cite{mansoorzare2023}, and phononic oscillators \cite{Wendt2026}, they are difficult to extend beyond these simple geometries. The homogeneous model, while admitting an exact solution, ignores the tensorial nature of the material so as to focus on a single polarization of plane waves; meanwhile, the perturbation-based model makes many approximations and cannot capture any elastic coupling that occurs when the semiconductor is in direct contact with the piezoelectric substrate, which is the most common scenario in real state-of-the-art devices. This highlights the limitations on calculating the exact acoustoelectric effects for real-world systems, for which the canonical models fail to maintain their accuracy.

\begin{figure}
\includegraphics[width=0.95\columnwidth]{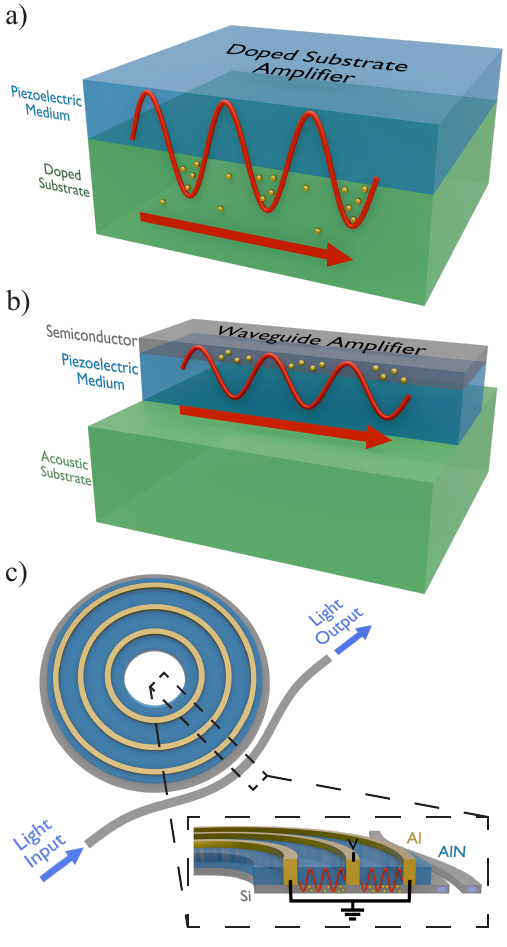}
\caption{a) Platform with piezoelectric material on top of a substrate that has been doped to have carriers available for the acoustoelectric effect. This structure is not covered by the canonical models due elastic coupling between the semiconductor and piezoelectric layers. b) Piezoelectric waveguides on a substrate with the semiconductor on top. The canonical models do not take into account rapid transverse horizontal variations in field strength introduced by the acoustic waveguide. c) Optomechanical cavity utilizing the acoustoelectric effect to alter acoustic breathing modes to control phononic properties for enhancement of Brillouin scattering \cite{Mack2024}. Such an arbitrary geometry is not covered by existing models.} 
\label{Fig:Arb}
\end{figure}

More comprehensive and arbitrary models are needed to accurately predict device performance in novel geometries including, but not limited to, thin piezoelectric films placed on top of a semiconductor substrate, spatially dependent carrier concentration in the semiconductor, phononic waveguides with strong horizontal and vertical confinement, and phononic cavities, a subset of which are illustrated in Figure \ref{Fig:Arb}. In particular, complicated geometries, such as the acoustoelectrically enhanced optomechanical cavity shown in Figure \ref{Fig:Arb}c, have no analog in the canonical models \cite{Mack2024}. Extending acoustoelectric models to these geometries is vital to optimizing acoustoelectric interactions, understanding electrical boundary conditions, controlling dispersion, and refining device architecture. It would be expected that the field of acoustoelectric devices would progress much faster if powerful and automatable tools, such as finite element method (FEM) modeling, were amenable to device design. As such, we develop and evaluate an FEM model that incorporates acoustoelectric interactions to solve for the acoustic modes and their corresponding voltage-dependent gain and dispersion curves in the presence of a DC-bias drift field.

Despite the ubiquity of FEM modeling for RF elastic waves in solids and semiconductor research, to the best of our awareness, there is nothing in the literature that describes an acoustoelectric multiphysics formulation of an FEM model that allows this problem to be solved. This is because commercially available FEM models do not provide the necessary couplings between piezoelectric waves and semiconductor electronics. We utilize a custom multiphysics coupling with mathematical modifications to the constraint equations, developed using COMSOL Multiphysics, to implement our acoustoelectric FEM model. The FEM model is used to solve for the eigenvectors and their corresponding complex eigenfrequencies in this non-Hermitian formulation of acoustoelectrics. Complex eigenfrequencies arise from the coupled non-Hermitian governing equations in the quasistatic regime, allowing us to extract loss and gain from the imaginary part of the eigenfrequency. \cite{Ashida2020,Kim2025,Xiao2025}.  

We start by detailing this coupling, which incorporates the full 3D drift-diffusion equations and the full 3D tensorial elastic and electromagnetic field equations in the quasistatic limit \cite{auld1973}. From there, we validate the FEM model by numerically comparing their predicted results to analytical results from the canonical models. After verifying the FEM model is in agreement with the canonical models, we conclude by evaluating the acoustoelectric effect in piezoelectric-on-semiconductor and phononic waveguide geometries, as illustrated in Figures \ref{Fig:Arb}a and \ref{Fig:Arb}b, respectively. These geometries are not supported by the canonical models, and we illustrate this by showing the canonical models cannot recreate the gain curves without modifying the material parameters. We demonstrate that the values of these modified parameters have no apparent correlation to the physical system and are sensitive to the exact geometry and material platform, making them extremely difficult to predict without using the FEM model to solve for the full and exact solution.

\section{The Acoustoelectric Interaction}

We start by developing the acoustoelectric governing equations in a manner that is conducive to implementation into an FEM model. The set of governing equations for an acoustoelectric mode in an arbitrary geometry is
\begin{equation}
    \mathbf{\nabla}\cdot\mathbf{D} = -q n_f,
    \label{Gauss}
\end{equation}
\begin{equation}
    \mathbf{J} = qn\mu\mathbf{E}+q\mathcal{D}\nabla{n},
    \label{CurrentDensity}
\end{equation}
\begin{equation}
    \mathbf{\nabla}\cdot\mathbf{J}=q\partial n_f/\partial t,
    \label{CurrentContinuity}
\end{equation}
\begin{equation}
    \mathbf{\nabla}\cdot\mathbf{T}=\rho\frac{\partial^2\mathbf{u}}{\partial t^2},
    \label{Acoustic Wave Equation}
\end{equation}
\begin{equation}
    \mathbf{T} = -\mathbf{e}\cdot\mathbf{E} + \mathbf{c}^E:\mathbf{S},
    \label{Stress}
\end{equation}
and
\begin{equation}
    \mathbf{D} = \epsilon^S\cdot\mathbf{E} + \mathbf{e}:\mathbf{S};
    \label{ElectricDisplacement}
\end{equation}
where $\mathbf{E}$ is the electric field, $\mathbf{D}$ is the electric displacement field, $\mathbf{J}$ is the electric current density, $\mathbf{u}$ is the displacement field, $\mathbf{T}$ is the stress tensor, $\mathbf{S}$ is the strain tensor, $q$ is the elementary charge, $\mu$ is the carrier mobility, $\epsilon^S$ is the dielectric tensor under constant strain, $\mathbf{e}$ is the piezoelectric coupling tensor, $\mathbf{c}^E$ is the elasticity tensor under constant electric field, and $\rho$ is the density of the material. $\mathcal{D}=\mu k_B T/q$ is the diffusion constant obtained from the Einstein relation for diffusion for charged particles. $n=n_0+f_tn_f$ is the carrier density, where $n_0$ is the equilibrium carrier density, $n_f$ is the carrier density contributing to $\mathbf{D}$, and $f_t$ is a trapping parameter to capture the behavior of carrier traps.

The trapping parameter, $f_t$,  is defined as 
\begin{equation}
    f_t = \frac{f_r-i\omega\tau}{1-i\omega\tau},
\end{equation}
where $\tau$ is the characteristic trapping time, $f_r$ is the ratio of the time that carriers spend in traps, and $\omega$ is the frequency of the acoustic wave \cite{Conwell1971}. We include $f_t$ for completeness and to illustrate the robustness of our model when comparing to the canonical models.

This formulation only utilizes the irrotational parts of the electric field, greatly simplifying the solutions to the system. These approximations are extremely accurate because the contributions from the rotational part of the field are of order $v_a/c_0 \sim10^{-5}$, where $v_a$ is the speed of sound in a material and $c_0$ is the speed of light \cite{auld1973}.

Assuming our fields are at a single frequency and contain a DC-bias drift field, $\mathbf{E}_0$, our electric field takes the form of $\mathbf{E}(\mathbf{r},t) = \mathbf{E}_0(\mathbf{r}) + \frac{1}{2}\left(\mathbf{E}_{\omega}(r)e^{i\omega t} + \mathbf{E}^*_{\omega}(\mathbf{r})e^{-i\omega t}\right)$ with $\mathbf{D}$, $\mathbf{J}$, $\mathbf{S}$, and $\mathbf{T}$ following equivalent expansions \cite{Conwell1971}.  Combining Equations \ref{Gauss}, \ref{CurrentDensity}, and \ref{CurrentContinuity} gives us an equation for the current density in the frequency domain,
\begin{equation}
    \mathbf{\nabla}\cdot\left(\mathbf{J}_{eff} + f_t\mathbf{J}_{AE}\right)=0,
    \label{ElectronicEquation_Frequency}
\end{equation}
where
\begin{equation}
    \mathbf{J}_{eff} = i\omega\mathbf{D}_\omega +\sigma_0\mathbf{E}_\omega,
    \label{J_eff}
\end{equation}
and
\begin{equation}
    \mathbf{J}_{AE} = \mathcal{D}\nabla\nabla\cdot\mathbf{D}_\omega+\mu\mathbf{E}_0\nabla\cdot\mathbf{D}_\omega+\mu\mathbf{E}_\omega\nabla\cdot\mathbf{D}_0.
    \label{J_AE}
\end{equation}

Using Equation \ref{ElectronicEquation_Frequency} to account for the drift-diffusion equations allows us to incorporate the acoustoelectric current density, $\mathbf{J}_{AE}$, from Equation \ref{J_AE} into our FEM model. To account for the higher order spatial derivative in the diffusion term of Equation \ref{J_AE} we add a dependent variable equal to $\nabla\cdot\mathbf{D}$. Mathematically, this is equivalent to breaking a higher order ordinary differential equation (ODE) into several smaller order ODEs to make it computationally tractable.

If we set $\mathbf{J}_{AE}=0$, Equation \ref{ElectronicEquation_Frequency} can be simplified down to Gauss's law such that
\begin{equation}
    \nabla\cdot\left(\mathbf{\epsilon}'\cdot\mathbf{E}+\mathbf{e}:\mathbf{S}\right)=0,
\end{equation}
where $\mathbf{\epsilon}'=\mathbf{\epsilon}^S-i(\sigma_0/\omega)\mathbf{I}$ and $\mathbf{I}$ denotes identity. From this we see that including the drift-diffusion equations into the governing equations of the acoustic modes makes our effective permittivity for the mode complex, resulting in loss when no DC drift field or free carriers are present. When $\mathbf{J}_{AE}\neq0$ and we apply a DC-bias drift to the semiconductor, our effective permittivity is now dependent on $\mathbf{E}_0$, resulting in a voltage-dependent loss or gain as is characteristic of the acoustoelectric effect.

This makes our system non-Hermitian, resulting in complex wavenumbers, $k=k_R+ik_I$, and complex eigenfrequencies, $\omega = \omega_R+i\omega_I$. Assuming time harmonicity and plane wave propagation in the $x$ direction, we seek modes of the form
\begin{equation}
    \mathbf{u}(\mathbf{r},t) = \mathbf{u}(y,z)e^{-ik_Rx}e^{i\omega_R t}e^{k_Ix-\omega_I t},
\end{equation}
where $k_I$ is interpreted as spatial loss or gain, depending on the sign. Likewise, $\omega_I$ corresponds to temporal loss or gain. 

To better understand the nature of $k_I$ and $\omega_I$ in a non-Hermitian system and how we can relate the spatial loss or gain to the temporal loss or gain found from an eigenfrequency solver for a spatially periodic system, we start with an arbitrary dispersion relation
\begin{equation}
    \omega = F(k),
    \label{Dispersion}
\end{equation}
where $k$ and $\omega$ may both be complex and $F$ is an analytic function near the mode of interest.

Taking $k=k_0+\delta k$ and $\omega=\omega_0+\delta\omega$ such that $\omega_0=F(k_0)$, we linearize $F(k)$ about $k_0$ such that
\begin{equation}
    \omega_0 +\delta\omega = F(k_0) + \delta k\left.\frac{dF(k)}{dk}\right|_{k=k_0}.
    \label{DispersionExpanded}
\end{equation}
Since $\omega_0 = F(k_0)$ by assumption and $dF(k)/dk = d\omega/dk\equiv v_g$, where $v_g$ is the group velocity, we can rewrite Equation \ref{DispersionExpanded} as
\begin{equation}
    \delta k  = \frac{\delta\omega}{v_g}.
    \label{ImagConv}
\end{equation}

If we limit $\delta k$ and $\delta \omega$ to be purely imaginary, Equation \ref{ImagConv} allows us to freely convert between spatial and temporal gain or loss and still satisfy Equation \ref{Dispersion}. This allows us to set up our FEM model with Floquet-Bloch boundary conditions such that $k$ is purely real. In order to satisfy the non-Hermitian dispersion relation, this forces $\omega$ to be complex. We can then use Equation \ref{ImagConv} to convert the temporal loss or gain to spatial loss or gain. This requires taking $\delta\omega = -\omega_I$, taking $\delta k = k_I$, and defining the acoustoelectric gain coefficient to be $\alpha = k_I$ such that
\begin{equation}
    \alpha = -\frac{\omega_I}{v_g}.
\end{equation}

By implementing the governing equations for an acoustoelectric mode, enforcing Equation \ref{ElectronicEquation_Frequency}, and  defining $\alpha$ as a spatial loss or gain term, we can now easily compare to the canonical models utilizing FEM computational techniques and extend our analysis to arbitrary geometries.

\begin{figure*}
\includegraphics[width=\textwidth]{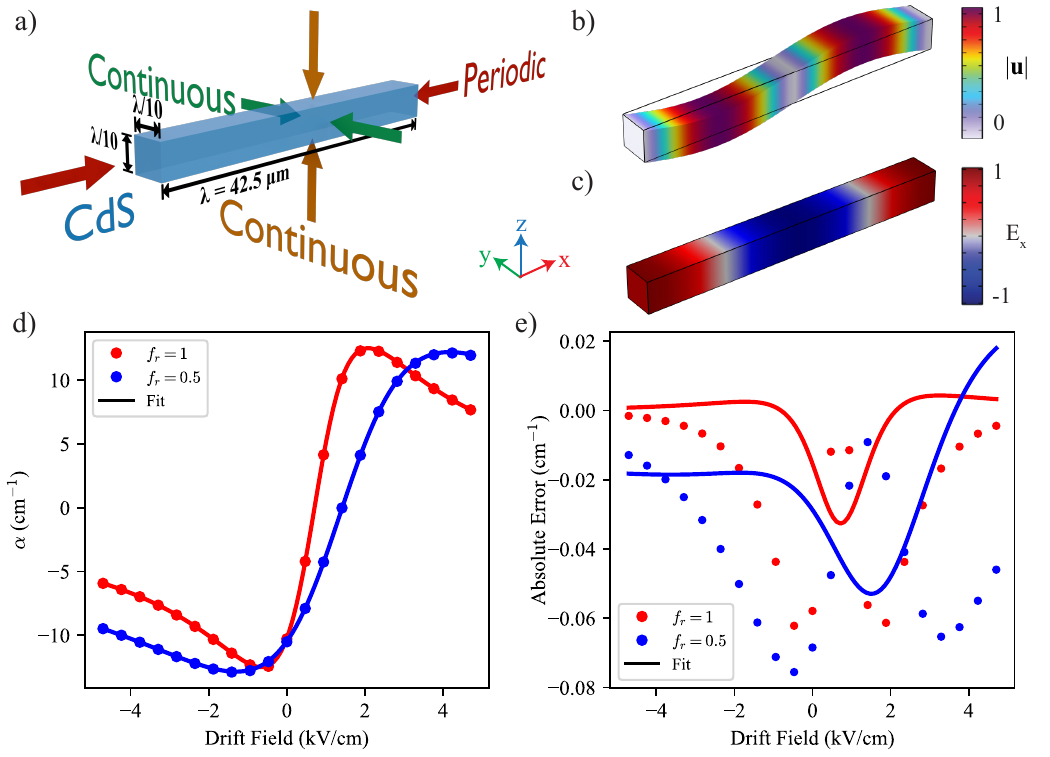}
\caption{a) Geometry of the CdS used in the simulation. We find a shear vertical plane wave solved for with a wavelength of $42.5$ microns and a frequency of $42$ MHz. b) Magnitude of the displacement, $|\mathbf{u}|$, for the plane wave. c) $x$-component of the electric field, $E_x$, for the plane wave. d) Gain curves as a function of applied DC electric field with and without trapping. Fits are obtained with the extracted parameters in agreement with the input parameters. e) Absolute error between the simulation and analytical model shown by the discrete points and absolute error between the fit and the analytical model shown by the continuous curves.} 
\label{Fig:Ganguly}
\end{figure*}

\section{Plane Waves in a Piezoelectric Semiconductor}

Our next task is to compare our FEM model to the two canonical models. We start with a bulk piezoelectric semiconductor, Cadmium Sulfide (CdS), one of the most widely studied piezoelectric semiconductors for acoustoelectric applications \cite{Conwell1971,Hutson1961,Uchida1964,Moore1965}. In this section, we compare our FEM results to analytical expressions for the DC-biased amplification of a bulk plane wave in wurtzite CdS. 

The DC-biased gain coefficient for a bulk plane wave in a volumetric acoustoelectric amplifier is derived as \cite{Conwell1971} 

\begin{widetext}

\begin{equation}
    \alpha = -\frac{e^2}{2\epsilon c}\frac{\omega_c}{v_a} \frac{1 - \operatorname{Re}(f_t)\mu E_0/v_a - \operatorname{Im}(f_t)\omega_m/\omega_D}{[1-\operatorname{Re}(f_t)\mu E_0/v_a - \operatorname{Im}(f_t)\omega_m/\omega_D]^2+[\omega_c/\omega_m - \operatorname{Im}(f_t)\mu E_0/v_a + \operatorname{Re}(f_t)\omega_m/\omega_D]^2},
    \label{AEGain}
\end{equation}
\end{widetext}
where $e$ is the effective piezoelectric coupling constant, $\epsilon$ is the dielectric constant, $c$ is the effective elasticity constant, $\omega_c=\sigma_0/\epsilon$ is the conductivity relaxation frequency, $\omega_d=v_a^2/\mathcal{D}$ is the diffusion frequency, $E_0$ is the applied DC-bias drift field. 

\begin{table}[b]
\caption{\label{tab:noTrap}%
Table comparing the input model parameters to the extracted parameters for $\omega_c$, $\mathcal{D}$, and $f_r$ utilizing a least squares fit for the no trapping case. Uncertainties are the standard deviations obtained from the least squares fit. Extracted parameters are seen to be in close agreement to the input parameters.
}
\begin{ruledtabular}
\begin{tabular}{cccc}
\textrm{Parameter}&
\textrm{Input}&
\textrm{Fit}\\
\colrule
$\omega_c \left(\text{rad/s}\right)$ & $5\times10^8$ & $\left(4.988\pm0.010\right)\times10^8$\\
$\mathcal{D} \left(\text{cm}^2\text{/s}\right)$ & $6.463\times10^{-4}$ & $\left(6.448\pm0.195\right)\times10^{-4}$\\
$f_r$ & 1 & $0.997\pm0.001$\\
\end{tabular}
\end{ruledtabular}
\end{table}

\begin{table}[b]
\caption{\label{tab:Trap}%
Table comparing the input model parameters to the extracted parameters for $\omega_c$, $\mathcal{D}$, and $f_r$ utilizing a least squares fit for the trapping case. Uncertainties are the standard deviations obtained from the least squares fit. Extracted parameters are seen to be in close agreement to the input parameters.
}
\begin{ruledtabular}
\begin{tabular}{cccc}
\textrm{Parameter}&
\textrm{Input}&
\textrm{Fit}\\
\colrule
$\omega_c \left(\text{rad/s}\right)$ & $5\times10^8$ & $\left(4.995\pm0.013\right)\times10^8$\\
$\mathcal{D} \left(\text{cm}^2\text{/s}\right)$ & $6.463\times10^{-4}$ & $\left(5.869\pm0.401\right)\times10^{-4}$\\
$f_r$ & 0.5 & $0.4980\pm0.0008$\\
\end{tabular}
\end{ruledtabular}
\end{table}

Utilizing the previously described coupling between the acoustic and electric fields, including the acoustoelectric current density from Equation \ref{J_AE}, we construct an FEM model with geometry and boundary conditions as shown in Figure \ref{Fig:Ganguly}a. Since we are only concerned with plane waves with single polarizations, we set continuous boundary conditions on the sides transverse to the direction of propagation and utilize Floquet-Bloch periodicity in the direction of propagation along the $x$-axis. The magnitude of displacement, $|\mathbf{u}|$, of the mode investigated is shown in Figure \ref{Fig:Ganguly}b and the longitudinal electric field, $E_x$, is shown in Figure \ref{Fig:Ganguly}c. As we are solving for plane waves in the basal plane of CdS, we desire a shear vertical plane wave such that we obtain a mode with only a longitudinal electric field, allowing us to suppress the tensorial nature of the piezoelectric tensor despite using the full tensorial representation of CdS in the FEM model. Suppressing the tensorial nature of the tensor elements allows us to easily make a direct comparison to Equation \ref{AEGain}.

Simulation results for the gain of a shear vertical plane wave in CdS are shown in Figure \ref{Fig:Ganguly}d.  Results are shown with and without trapping and fits to equation \ref{AEGain} are included for both cases. The input parameters and extracted fit parameters for the no trapping case are shown in Table \ref{tab:noTrap} while the same information for the trapping case is included in Table \ref{tab:Trap}. The fit parameters are obtained via a least squares fit and the uncertainty is taken to be the standard deviation obtained from the covariance matrix of the fit parameters. The fit parameters are in close agreement with the input model parameters, indicating our FEM model is working correctly. To further verify the model and fit performance when compared to the analytical model, an absolute error plot comparing simulation results and fits to Equation \ref{AEGain} is shown in Figure \ref{Fig:Ganguly}e. The discrete points correspond to the simulation error while the solid line indicates the curve fit error when compared the analytical model. From this, we see the absolute error is extremely small for all simulation results, indicating the FEM model is in agreement with the first canonical model for the case of a bulk acoustic plane wave traveling in a piezoelectric semiconductor.

\section{Rayleigh Waves in the Separated Medium Amplifier}

We now turn our attention to the case of a separated medium Rayleigh wave amplifier. This system has been the subject of focus both experimentally and theoretically \cite{Hackett2021}. In this section, we focus on comparing our FEM model to a perturbative model utilizing a normal mode expansion of the SAWs. This perturbation theory was developed by understanding how the field admittance changes in the presence of a semiconductor that is separated from the acoustic region by an air gap. An example of this system is included in Figure \ref{Fig:Intro}b. This model does not fully account for the system where the air gap goes to zero ($h=0$) as it does not incorporate elastic couplings across the semiconductor-piezoelectric boundary.

\begin{figure*}
\includegraphics[width=\textwidth]{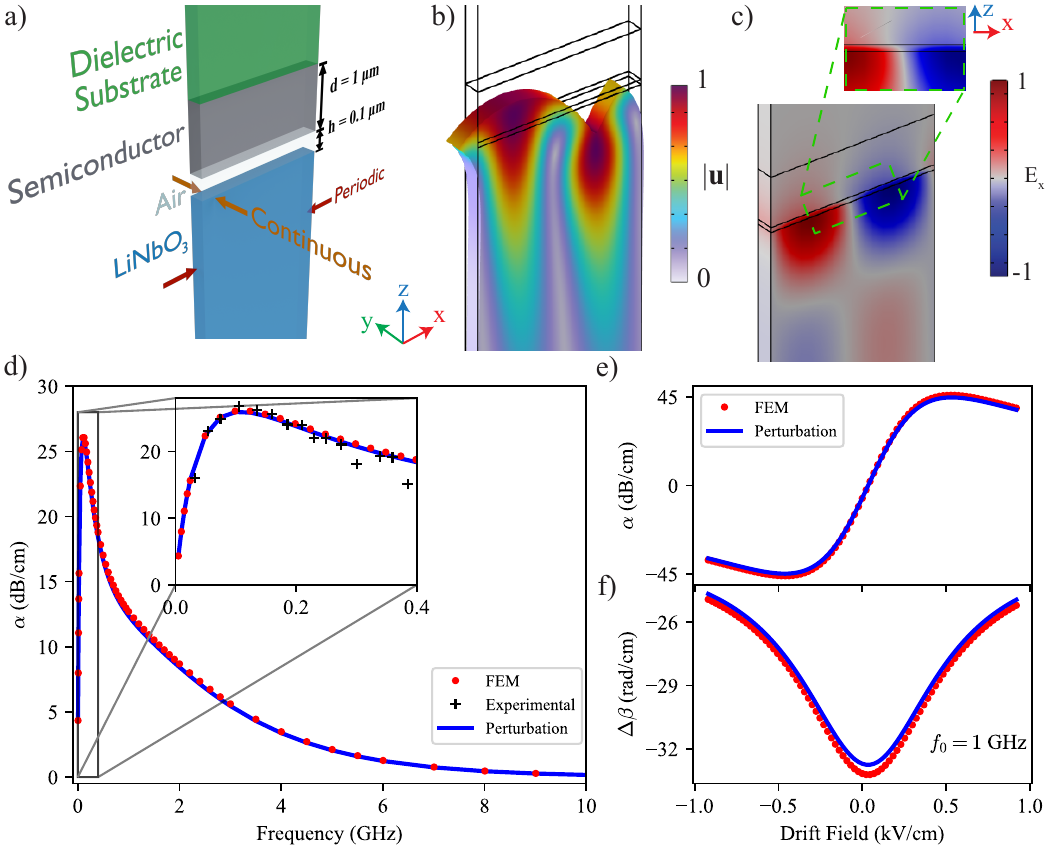}
\caption{a) Simulation geometry of the separated medium amplifier. b) Magnitude of the displacement, $|\mathbf{u}|$, for the Rayleigh mode. c) $x$-component of the electric field, $E_x$, for the Rayleigh mode. Inset image shows $E_x$ edge on with the surface parallel to $xz$-plane to better illustrate the phase shift in the electric field caused by screening effects. d) Plot of acoustoelectric gain, $\alpha$, as a function of frequency with electron drift velocity fixed at $v_d=3v_a$. Inset shows a portion of the curve corresponding to experimental data from the original work used to verify the perturbation-based model \cite{Kino1971}. Experimental data points have been estimated utilizing digital extraction tools. e) Plot of $\alpha$ as a function of drift field at $1$ GHz. f) Plot of change in wavenumber, $\Delta\beta$, as a function of drift field at $1$ GHz} 
\label{Fig:KinoReeder}
\end{figure*}

In the original paper, the equations for the gain and dispersion were approximated from more general expressions to bring the final equations into a similar form to the volumetric acoustoelectric gain\cite{Kino1971}. Here we derive the more accurate equations from earlier in the paper to make a better comparison to our FEM model. As such, we start with
\begin{equation}
    \frac{1}{Y^--Y_0} = i\left[\frac{M(\beta h)}{\beta\omega\epsilon_0}-\frac{\omega\beta_a w Z_a(\beta h)}{\beta^2-\beta_a^2}\right],
    \label{PiezoCoupling}
\end{equation}
which is referred to as the piezoelectric coupling equation in the original work, where $Y^-$ is the admittance at the bottom edge of the semiconductor located at $z=h$, $Y_0=i\omega\beta\epsilon_0$ is the admittance at $z=h$ without the semiconductor present, $\beta_a$ is the unperturbed wavenumber, $\beta$ is the perturbed wavenumber, $w$ is the width of the amplifier, and $\epsilon_0$ is the free space permittivity. The expression for $M(\beta h)$, which is termed the space charge potential factor, is given as
\begin{equation}
    M(\beta h) = \frac{\epsilon_0+\epsilon_p\tanh{(\beta h)}}{(\epsilon_0+\epsilon_p)(1+\tanh{(\beta h)})},
    \label{M}
\end{equation}
where $\epsilon_p$ is the effective dielectric permittivity of the piezoelectric material. The interaction impedance, $Z_a(\beta h)$, is given as 
\begin{equation}
    Z_a(\beta h) = \frac{\phi_a(h)\phi_a^*(h)}{2P_a}
    \label{Z}
\end{equation} at $z=h$. $\phi_a$ is the electric potential associated with the acoustic mode and $P_a$ is the total power propagating in the unperturbed surface acoustic wave (SAW) mode.

Taking the perturbed wavenumber to be of the form 
\begin{equation}
    \beta=\beta_a+\Delta\beta+i\alpha,
\end{equation}
and assuming the perturbations are small in comparison to $\beta$ such that $\alpha,\Delta\beta<<\beta$ we find $\beta^2\approx\beta_a^2+2\beta\Delta\beta+2i\alpha\beta$ and Equation \ref{PiezoCoupling} can be rearranged to get the most general equations for the gain and dispersion
\begin{equation}
    \alpha = \frac{1}{2}\operatorname{Im}\left[\frac{w\beta Z_a(\beta h)}{i/(Y^--Y_0)+M(\beta h)/\beta\omega\epsilon_0}\right],
    \label{KR_gain}
\end{equation}
and
\begin{equation}
    \Delta\beta = \frac{1}{2}\operatorname{Re}\left[\frac{w\beta Z_a(\beta h)}{i/(Y^--Y_0)+M(\beta h)/\beta\omega\epsilon_0}\right].
    \label{KR_DeltaBeta}
\end{equation}
In the original work, Equations \ref{KR_gain} and \ref{KR_DeltaBeta} are simplified by assuming the carrier response is much faster than the frequency of the acoustic wave to bring them into a form that more closely resembles the expression for the volumetric acoustoelectric gain, given in Equation \ref{AEGain} for $f_t=1$. To compare across a broad frequency range, we avoid making this assumption.

We compare Equations \ref{KR_gain} and \ref{KR_DeltaBeta} for a finite-thickness semiconductor to our FEM model in Figure \ref{Fig:KinoReeder}. $Y^-$ for a finite-thickness semiconductor has been previously derived and is included in the appendix for completeness, as well as $Y^-$ for the infinitesimally thin and semi-infinitely thick cases \cite{Kino1971}. The structure of the separated medium model is shown in Figure \ref{Fig:KinoReeder}a, where we utilize $Y$ cut, $Z$ propagating lithium niobate (LiNbO$_3$). The mode displacement, $|\mathbf{u}|$, and the longitudinal electric field, $E_x$, for the Rayleigh mode  are shown in Figures \ref{Fig:KinoReeder}b and \ref{Fig:KinoReeder}c, respectively. It is worth pointing out that we can observe screening effects on the electric field as evidenced by the phase shift of the field above the piezoelectric layer induced by the free carriers within the semiconductor layer as seen in the inset image of Figure \ref{Fig:KinoReeder}c which is taken edge on with the surface parallel to the $xz$-plane.

Sweeping the drift field for a Rayleigh mode at 1 GHz, our model is in close agreement to the gain curve obtained from Equation \ref{KR_gain}, as shown in Figure \ref{Fig:KinoReeder}e. In Figure \ref{Fig:KinoReeder}f, for the same mode, we demonstrate the shift in the propagation constant is also in close agreement between our FEM model and Equation \ref{KR_DeltaBeta}. For a fixed drift velocity of $v_d=3v_a$, we show the gain from our FEM model closely follows the results obtained from Equation \ref{KR_gain} over a broad frequency range spanning from 20 MHz to 10 GHz in Figure \ref{Fig:KinoReeder}d. The inset image shows the regime of comparison from the original work and includes experimental data points extracted from the original paper used to initially verify the model \cite{Kino1971}. From this, we conclude that our FEM model is in agreement with the second canonical model for the case of a separated medium amplifier.

\begin{figure}
\includegraphics[width=0.95\columnwidth]{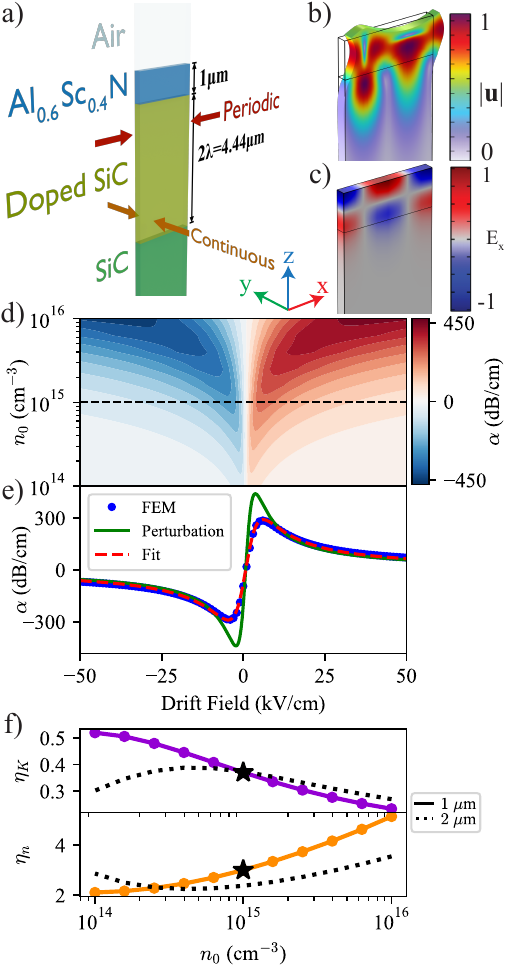}
\caption{a) Diagram of the simulation geometry for 1 micron thick Al$_{0.6}$Sc$_{0.4}$N grown on 4H SiC. The top layer of SiC with a thickness of $2\lambda$ is doped with a uniform carrier concentration and acts as the semiconductor for acoustoelectric interactions. b) Displacement magnitude, $|\mathbf{u}|$, of the Sezawa mode. c) $x$-component of Electric field, $E_x$, for the same Sezawa mode. d) Filled contour plot showing the gain, $\alpha$, as a function of the carrier concentration, $n_0$, and the applied drift field. e) $\alpha$ as a function of the applied drift field for $n_0=1\times10^{15}$ cm$^{-3}$. An initial estimate and a fit to the perturbation-based model are included. Location on the contour plot is denoted by the dashed line in d. f) Correction ratios, $\eta_K=K^2_{eff}/K^2$ and $\eta_n=n_{eff}/n_0$, depicting the difference between the fit parameters and expected initial estimates. The star indicates the case plotted in e. The dashed lines are the same correction ratios for 2 micron thick film of AlScN.} 
\label{Fig:DopedResults}
\end{figure}

\section{Sezawa Waves in Piezoelectric-on-Semiconductor Heterostructures}
At this point, we have verified that our FEM model has recreated the behavior of the two canonical models. Now, for the first time, we will present results for a geometry not captured by the canonical models. Here, we look at a piezoelectric material placed directly above a doped, high-velocity semiconductor substrate. Recent work has investigated this system and utilized the perturbative canonical model with modified parameters to analyze the experimental results \cite{hackett2024s}. Without more complete models of the acoustoelectric effect, such as our FEM model, it is impossible to say whether these modified parameters arise from nonidealities in the device or if the model does not predict the appropriate response for this system. 

A diagram of this piezoelectric-on-semiconductor heterostructure is shown in Figure \ref{Fig:DopedResults}a. We will be investigating the Sezawa mode in a 1 micron scandium doped aluminum nitride (Al$_{0.6}$Sc$_{0.4}$N) placed on top of a 4H Silicon Carbide (SiC) substrate with a uniform doping layer of $2\lambda=4.44\text{ }\mu$m. The Sezawa mode, shown in Figures \ref{Fig:DopedResults}b and \ref{Fig:DopedResults}c, is the first higher order Rayleigh mode found in slow-on-fast heterostructures \cite{HadjLarbi2019}. We focus on the Sezawa mode due to its large electromechanical coupling and emphasis in recent work as a promising mode for phononic circuits \cite{Deng2025,Du2024}. The Sezawa mode is highly dissimilar from a Rayleigh mode since it has much more strain localized to the piezoelectric-substrate interface, which leads to increased acoustoelectric interactions when compared to the Rayleigh mode in a piezoelectric-on-semiconductor heterostructure. This makes the Sezawa mode a promising candidate for acoustoelectric amplifiers in the thin film piezoelectric-on-semiconductor system.

We run several simulations with varying carrier concentrations and DC drift fields as shown in Figure \ref{Fig:DopedResults}d. In Figure \ref{Fig:DopedResults}e, we compare the gain curve for $n_0=1\times10^{15}$ cm$^{-3}$ to an approximation based on Equation \ref{KR_gain} and used in recent experimental work \cite{hackett2024s}. In this approximation, we utilize the interaction impedance at the boundary of the AlScN and SiC and swap the free space permittivity with the permittivity of the SiC. This approximation results in an overestimate of $\alpha$ and maximum gain is achieved at a lower drift field.

To understand if these failings can be remedied by changing the effective interaction impedance $Z_{eff}$ and carrier concentration $n_{eff}$, we fit our data to Equation \ref{KR_gain} with these taken as free parameters. The interaction impedance is proportional to the electromechanical coupling, $Z_a\propto K^2$ \cite{Kino1971}, so by fitting $Z_{eff}$, we are essentially correcting the electromechanical coupling $K_{eff}^2$. It is worthwhile to note that we can find an acoustoelectric gain curve that fits the data well, but with unintuitive values of $K^2_{eff}$ and $n_{eff}$. The correction ratios, $\eta_K = K^2_{eff}/K^2$ and $\eta_n=n_{eff}/n_0$, for these free parameters are shown in Figure \ref{Fig:DopedResults}f and exhibit a nontrivial dependence on $n_0$. The dashed line indicates the $\eta_K$ and $\eta_n$ for a 2 micron thick layer of AlScN. This illustrates the requisite modification of $K^2_{eff}$ and $n_{eff}$ to predict accurate device behavior is difficult to determine based on geometrical parameters alone, but rather requires an understanding of how the modes interact with and change in the presence of the carriers, which greatly negates the simplicity and utility of the model. The difference between the correction ratios for the 1 micron and 2 micron thick cases illustrate that these correction ratios are highly geometry-dependent. The existing canonical models do not account for this geometry-dependent nature of the acoustoelectric effect and the appropriate corrections are difficult to predict without utilizing a full FEM solver that accounts for acoustoelectric interactions in anisotropic media and heterogeneous material platforms.

\begin{figure}
    \includegraphics[width=0.95\columnwidth]{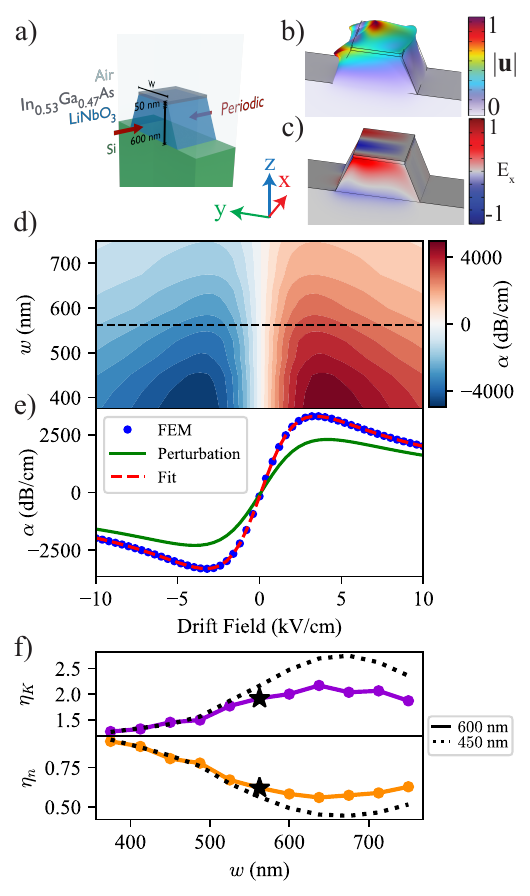}
    \caption{a) Diagram of geometry utilized for an acoustoelectric simulation in a waveguide geometry built in LiNbO$_3$ on Si with a sidewall angle of 78 degrees and thickness of 600 nm. LiNbO$_3$ is orientated so that the material geometry is Y-cut and the mode is X-propagating. b) Displacement magnitude, $|\mathbf{u}|$, of the shear horizontal mode.  c) $x$-component of electric field, $E_x$, for the shear horizontal mode. d) Filled contour plot showing the gain, $\alpha$, as a function of the waveguide width, $w$, and drift field. e) $\alpha$ as a function of the drift field for $w=562.5$ nm. An initial estimate and a fit to the perturbation-based model are included. Location on the contour plot is denoted by the dashed line in d. f) Correction ratios, $\eta_K=K^2_{eff}/K^2$ and $\eta_n=n_{eff}/n_0$, depicting the difference between the fit parameters and initial estimates. The star indicates the case plotted in e. The dashed lines are the same correction ratios for a waveguide with a thickness of 450 nm.}
    \label{Fig:WGResults}
\end{figure}

\section{Shear Horizontal Waves in a Fully 2D Waveguide}
The existing canonical models make the simplification of utilizing the 1D drift-diffusion equations, only dealing with the variations in the carrier concentration in the direction of propagation for the acoustic wave. This makes it difficult to extend these models accurately to acoustic waveguides with large horizontal confinement, where the waveguide width is on the order of one wavelength. This problem is analogous to the difficulties of solving for closed form solutions to horizontally guided optical waveguides, but with increased complexity due to the tensorial nature of the acoustic wave equations. Since our FEM model utilizes the full 3D drift-diffusion equations, we are able to account for these horizontal variations without altering our model.

To understand how variation in the transverse horizontal direction of the field affects the calculated acoustoelectric gain, we turn to a waveguide geometry utilizing a ridge waveguide in Y-cut, X-propagating lithium niobate (LiNbO$_3$) on silicon (Si) as shown in Figure \ref{Fig:WGResults}a. The asymmetry of the mode results form the strong anisotropy of the lithium niobate. We include a sidewall angle of 78 degrees take into account realistic fabrication constraints. A $50$ nm layer of indium gallium arsenide (In$_{0.53}$Ga$_{0.47}$As) is placed directly on top of the waveguide, providing access to carriers for the acoustic wave. We desire a fixed frequency of $f=4.7 GHz$ when the acoustoelectric effect is turned off in the FEM model. This is to reduce the parameters to be varied when comparing to the perturbative approximation and to provide a better comparison between different waveguide widths. As such, we vary the wavelength as a function of waveguide width, $w$, to account for the shift of the dispersion relation. This results in a span of wavelengths ranging from $\lambda=728.8$ nm at $w=375$ nm to $\lambda=762$ nm at $w=750$ nm. For these parameters we find a shear horizontal mode as shown in Figures \ref{Fig:WGResults}b and \ref{Fig:WGResults}c.

Varying the applied drift field across the In$_{0.53}$Ga$_{0.47}$As layer, we obtain gain curves for different waveguide widths, $w$, as shown in Figure \ref{Fig:WGResults}d. The gain curve for $w=562.5$ nm is shown in Figure \ref{Fig:WGResults}e. An approximation obtained from Equation \ref{KR_gain} and a fit are also included in this plot. Since the perturbative approach assumes no transverse horizontal variation in the field values of the acoustic wave, we modify Equation \ref{KR_gain} slightly by making the substitution 
\begin{equation}
    wZ_a\rightarrow\tilde{Z}_a=\int_{WG}Z_a dy,
\end{equation}
where $w$ is the waveguide width, and the integral is taken along the $y$-axis over the top of the waveguide at the interface between the LiNbO$_3$ and the In$_{0.53}$Ga$_{0.47}$As. This reduces to the original formula in the case of uniform transverse variation, suggesting it is a reasonable modification to the perturbative model.

Our fits are similar to the ones done previously for the Sezawa mode but instead using $\tilde{Z}_{eff}$ as a free parameter instead of $Z_{eff}$. $\tilde{Z}_{eff}$ is still proportional to $K^2$, \cite{Kino1971}, so we use the same correction ratios $\eta_K$ and $\eta_n$ to evaluate our fitted parameters to the initial estimates, as shown in Figure \ref{Fig:WGResults}f. We observe a nontrivial dependence on the waveguide width $w$. Similar to the AlScN on SiC case in the previous section, we see that these correction ratios change as we alter the thickness of the piezoelectric layer, highlighting the strong geometry dependence of the acoustoelectric effect. From these correction ratios, we see that we need to significantly modify our model parameters from their expected inputs to obtain accurate gain curves from the perturbative model. Without the full FEM solution, it is difficult to know what these modified parameters should be, highlighting the advantage of solving for the acoustic modes while accounting for the acoustoelectric effect.

\section{Conclusion}
Having compared our FEM model to the two canonical models used extensively in predicting the behavior of acoustoelectric devices, we see that our FEM model accurately predicts the amount of acoustoelectric gain seen for various acoustic modes in different material platforms. In Figure \ref{Fig:Ganguly}, we see that the volumetric acoustoelectric gain is accurately captured by our FEM model. Comparing to a perturbative approximation of acoustoelectric gain in the separated medium amplifier, we once again see agreement between our FEM model and a prevalent model for acoustoelectric gain as seen in Figure \ref{Fig:KinoReeder}.

From there, we illustrate that acoustoelectric gain curves can be obtained for more arbitrary geometries that are not covered by existing models. We start with the case where the piezoelectric material is placed directly on top of a semiconductor substrate. In particular, we look at AlScN on top of doped 4H SiC and the results are shown in Figure \ref{Fig:DopedResults}. From these results, we show that modifying the canonical models to accurately capture this piezoelectric-on-semiconductor heterostructure is unintuitive and extremely difficult to predict accurately without utilizing a full FEM model. We arrive at a similar conclusion when looking at modes in a fully 2D waveguide with large transverse horizontal variation in the field profiles as seen in Figure \ref{Fig:WGResults}.

From this discussion, we see that an FEM model which can be utilized to study the acoustoelectric effect in arbitrary geometries is a powerful tool that can provide insights towards future efforts in creating novel devices and exploring complex systems. Possible areas of interest include utilizing the acoustoelectric effect in cavity systems, modifying phonon dissipation rates in optomechanical systems, as well as accurately predicting effective nonlinearity coefficients for nonlinear acoustoelectric devices.

Acoustoelectric devices show great potential for a paradigm shift in RF signal processing and sensing with SAWs. For the most efficient devices, we desire modes with electromechanical coupling as large as possible. This can quickly push us out of the perturbative regime of acoustoelectrics, which is where the canonical models are accurate. To provide predictive power for models in this device regime, we need more accurate and complete modeling of the acoustoelectric effect that accounts for non-perturbative effects. For modes with large electromechanical coupling, it is expected--and we have observed in our model--that changing the carrier concentration and applying a drift field alters the modal properties of the acoustic waves. We will show in future works that our FEM model accounts for these non-perturbative effects. Furthermore, the ability to model arbitrary geometries, materials, and boundary conditions opens up a wide array of novel devices that have no analog in the canonical models or current acoustoelectric devices, including but not limited to acoustoelectric phononic crystals, exceptional point-based acoustoelectric devices, and acoustoelectric topological SAW devices.

\begin{figure}
    \includegraphics[width=0.95\columnwidth]{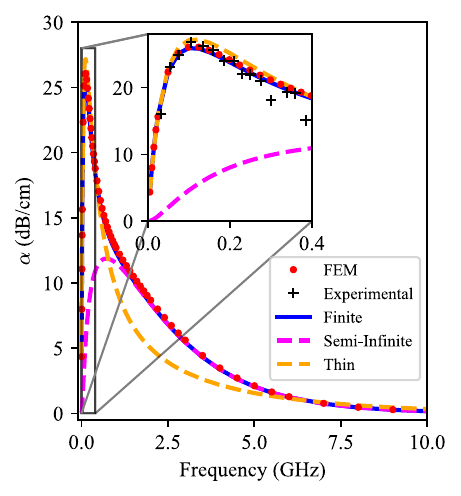}
    \caption{Same plot as Figure \ref{Fig:KinoReeder}d, but with the gain for the infinitesimally thin and semi-infinite semiconductor cases. We see the finite-thickness and FEM results come into agreement with the infinitesimally thin case for lower frequencies when the wavelength is long compared to the thickness of the semiconductor, $\lambda>>d$. Similarly the finite-thickness and FEM results come into agreement with the semi-infinite case for higher frequencies when the wavelength is very short compared to the thickness of the semiconductor, $\lambda<<d$. } 
    \label{Fig:KR_Supplement}
\end{figure}

\appendix*

\section{Admittances}
Here, we present the admittances for three semiconductor geometries for the separated medium amplifier. These have been previously derived and are included in this work for completeness \cite{Kino1971}.

For an infinitesimally thin semiconductor with a dielectric substrate, the admittance is found to be
\begin{equation}
    Y_{thin}^- = i\omega\beta\epsilon_d - \frac{\omega_c\epsilon_s\beta_a^2d}{(v_d/v_a-1)+i\omega_c/\omega_D},
    \label{thin_SI}
\end{equation}
where $\epsilon_d$ is the permittivity of the dielectric substrate the infinitesimally thin semiconductor is placed on.

For the semi-infinite case, the admittance is found to be
\begin{equation}
    Y^-_{inf} = j\omega\beta\epsilon_s + \frac{\omega\omega_c\epsilon_s\beta(\gamma-\beta)}{\gamma(\omega-\beta v_d)-j\omega_c\beta},
    \label{Inf_SI}
\end{equation}
where $\epsilon_s$ is the permittivity of the semiconductor, and $\gamma^2 = \left(\omega_c+i\left(\omega-\beta v_d\right)\right)/\mathcal{D} + \beta^2$ is the rate of screening of the electric field. For small frequencies, $\gamma = \sqrt{\omega_c\omega_D}/v_a = 1/\lambda_D$, where $\lambda_D$ is the Debye length We utilize the full form of $\gamma$ to limit the number of assumptions when comparing the results from our FEM model to the perturbation theory.

For the finite-thickness case, an assumption is made that the potential does not drop drastically over the width of the semiconductor to get the equation

\begin{equation}
    Y_{fin}^-=\frac{\omega-\beta v_d-i\omega_c}{(\omega-\beta v_d)/Y_L^- - i\omega_c/Y_D^-},
    \label{finite_SI}
\end{equation}
where $Y_L^-=i\beta\omega\epsilon_s\tanh{(\beta d)}$ and $Y_D^-=i\beta\omega\epsilon_s\tanh{(\gamma d)}$ are admittances that correspond to a solenoidal wave and a diffusion carrier wave in the semiconductor, respectively.

The gain obtained from Equations \ref{thin_SI}, \ref{Inf_SI}, and \ref{finite_SI} are plotted in Figure \ref{Fig:KR_Supplement} and compared to our FEM results. We see that at low frequencies, where $\lambda>>d$, $Y_{thin}^-$ predicts the gain accurately. Likewise at high frequencies, where $\lambda<<d$, $Y_{inf}^-$ predicts the gain accurately. For all frequency ranges the FEM model and $Y_{fin}^-$ are in agreement, indicating the finite-thickness expression for the admittance is best in general and corresponds most closely to real-world systems, as expected.


\bibliography{ref}

\end{document}